\documentclass[preprint,aps]{revtex4}
\usepackage{graphicx}%
\usepackage{dcolumn}
\usepackage{amsmath}
\usepackage{longtable}
\begin{document}
\title
{ Application of Van Der Waals Density Functionals to Two Dimensional Systems 
  Based on a Mixed Basis Approach }
\author
{Chung-Yuan Ren$^{a,\dagger}$, 
 Yia-Chung Chang$^{b,c}$, and Chen-Shiung Hsue$^{d}$  }
\affiliation
{ $^{a}$ Department of Physics, National Kaohsiung Normal University,
Kaohsiung 824, Taiwan  	\\
$^{b}$ Research Center for Applied Sciences, Academia Sinica, Taipei 115, Taiwan \\
$^{c}$ {Department of Physics, National Cheng-Kung University, Tainan 701, Taiwan}\\
$^{d}$ Department of Physics, National Tsing Hua University,
Hsinchu 300, Taiwan \\
$\dagger$ {\it E-mail address:} cyren@nknu.edu.tw }

\begin{abstract}
A van der Waals (vdW) density functional was implemented in the mixed basis 
approach previously developed for studying two dimensional systems, in which 
the vdW interaction plays an important role. 
The basis functions here are taken to be the localized B-splines 
for the finite non-periodic dimension and 
plane waves for the two periodic directions. 
This approach will 
significantly reduce the size of the basis set, especially for large systems, 
and therefore is computationally 
efficient for the diagonalization of the Kohn-Sham Hamiltonian.  
We applied the present algorithm to calculate the binding energy 
for the two-layer graphene case and the results are consistent with data 
reported earlier.    
We also found that, due to the relatively weak vdW interaction, 
the charge density obtained self-consistently for the whole bi-layer 
graphene system is not significantly different from the simple 
addition of those for the two individual one-layer system, except when the
interlayer separation is close enough that the  
strong electron-repulsion dominates. This finding suggests an efficient way to 
calculate the vdW interaction for large complex systems involving the 
Moir\'{e} pattern configurations.  \\
PACS: 71.15.Mb, 73.20.-r	\\
\end{abstract}
\maketitle
\section{INTRODUCTION}
The electronic properties of two-dimensional (2D) systems are fundamentally
different from those in higher dimensions due to their unusual collective
excitations. Among these 2D materials, graphite is the most well known. 
Graphite has a layered planar structure and 
is electrically conductive along the planes, whereas diamond, another allotrope
of carbon, is an insulator.  
Graphene is an isolated sheet of graphite and can be stacked via 
the weak van der Waals (vdW) interaction to form the graphite structure. 
There has been increasing interest in vdW graphene-based composite systems,
e.g., alkali metal/graphite adsorption systems \cite{BKLW,ZKSH} or 
MoS2/graphene heterostructures \cite{PHABNP}.   
Expectations concerning the creation  
of improved functional electrodevices with 
better performance characteristics are rising from 
the intensive exploration of such graphene-based materials.


First-principles methods based on the density functional theory 
within local density approximation (LDA) and 
generalized gradient approximation(GGA) have proven to
be powerful and successful in investigating static and dynamic 
properties of materials with strong ionic, covalent and metallic interactions. 
Unfortunately, these methods fail to describe the weak vdW 
dispersion interaction properly. 
For example, GGA calculations show no relevant binding 
between graphite sheets \cite{RJHSLL}.
While the LDA approach predicts an underestimated minimum for 
graphite \cite{LM}-\cite{NPA}, it cannot 
capture the vdW physics \cite{ZKSH}. 
Neither of these traditional functionals has basis to 
address issues of transferability for soft-matter problems involving the weak 
vdW bonding. To remedy this situation, a new approach using a van der Waals
density functional (vdW-DF) with a nonlocal correlation energy 
has been developed by Dion {\it et al} \cite{DRSLL}. 
This formalism accounts for the dominant dispersion energy, 
which is not correctly treated in standard DFT functionals. 

In this work, we implement the vdW-DF functionals 
using the mixed basis approach developed previously \cite{RHC,RCH}. 
The basis functions here are taken to be the localized B-splines 
for the finite non-periodic dimension and 2D
plane waves for the two periodic directions. 
B-splines are highly localized and piecewise polynomials 
\cite{deBoor}, which  
have proven to be an excellent tool for the description of 
wavefunctions in a real-space approach \cite{RHC}-\cite{RJH}. 
Such a mixed-basis method \cite{LC,LC1} avoids 
the use of  artificial vacuum layers of large thickness 
introduced by the supercell modeling,  
reducing  significantly the number of the basis functions, and   
therefore easing the computational burden for the 
diagonalization of the Kohn-Sham Hamiltonian. Another advantage of the 
present mixed basis method is that, 
for charged systems, the spurious Coulomb interaction between the defect, 
its images and the compensating background charge in the supercell approach can
be automatically avoided. No further modification needs to be made in the 
total-energy calculation \cite{RCH}.

We tested the present algorithm by studying the binding energy between 
two graphene sheets stacked in both AA and AB types, as depicted in  
Fig. \ref{fig1}. In addition, the charge density 
around each graphene sheet was found not to be significantly affected by the 
existence of another one, except when the two sheets are so close 
that the electron distributions of individual sheets overlap with each other and 
the strong electron repulsion dominates. 
This revelation would allow us to calculate the binding energy by the 
rigid-density model, i.e., the whole charge density for the system is simply 
the sum of those self-consistently calculated 
for the individual layers, instead of using the very 
time-consuming self-consistent density calculation for the whole system.  
The justified rigid-density model enables a simpler yet accurate evaluation of  
the vdW interaction for large complex systems including the 
Moir\'{e} pattern configurations.  
The results will be presented and discussed in details. 

%
\section{METHOD OF CALCULATION}
\subsection{B-splines}
For the sake of completeness, we
first briefly summarize the B-spline formalism.
More details can be found in Refs. \cite{RHC} and \cite{deBoor}.
In general, B-spline of order $\kappa$ consists
of positive polynomials of degree $\kappa-1$, over $\kappa$
adjacent intervals. These
polynomials vanish everywhere outside the subintervals $\tau_i < z <
\tau_{i+\kappa} $.
The B-spline basis set of order $\kappa$ with the knot sequence $\{ \tau_i\}$
is generated by the following relation :
\begin{equation}
B_{i,\kappa}(z)=\frac{z-\tau_i}{\tau_{i+\kappa-1}-\tau_i}B_{i,\kappa-1}(z)+\frac{\tau_{i+\kappa}-x}{\tau_{i+\kappa}
-\tau_{i+1}}B_{i+1,\kappa-1}(z),
\end{equation}
with
\begin{equation}
B_{i,1}(z)= \left \{ \begin{array}{ll}
                1,      & \tau_i \leq z < \tau_{i+1}  \\
                0,      & {\rm otherwise \ .}
                \end{array}
                \right.
\end{equation}
The first derivative of the B-spline of order $\kappa$ is given by
\begin{equation}
\frac{d}{dz}B_{i,\kappa}(z)=\frac{\kappa-1}{\tau_{i+\kappa-1}-\tau_i}B_{i,\kappa-1}(z)-\frac{\kappa-1}
{\tau_{i+\kappa}-\tau_{i+1}}B_{i+1,\kappa-1}(z).
\end{equation}

Therefore, the derivative of B-splines of order $\kappa$ is simply a 
linear combination of B-splines of order $\kappa-1$, which is
also a simple polynomial and is continuous across the knot sequence.
Obviously, B-splines are flexible to accurately represent any localized 
function of $z$ with a modest number of the basis by only
increasing the density of the knot
sequence where it varies rapidly.

\subsection{vdW-DF functional}
The nonlocal energy functional proposed  by 
Dion {\it et al.} \cite{DRSLL} is  
\begin{equation}
E^{vdW-DF}_{xc}= 
E^{revPBE}_{x}+
E^{LDA}_{c}+
E^{nl}_{c}.   \label{vdw-xc}
\end{equation}	
The first two parts are simply revPBE exchange \cite{ZY} and 
LDA correlation \cite{PW}.  
$E^{nl}_{c}$ is a non-local correlation functional 
that was introduced to accounts for dispersion interactions, and is  
given as 
\begin{equation}
E^{nl}_{c}= \frac{1}{2}  
\int 
\int 
{ d{\bf r_1} d{ \bf r_2} }
n({\bf r_1}) 
\phi(q_1,q_2,r_{12}) 
n({\bf r_2}), \label {eqnl}
\end{equation}
where $r_{12}=|\bf{r_1}-\bf{r_2}|$, and 
$q_1,q_2$ are the values of a universal function $q_0$ at 
$\bf{r_1}$ and $\bf{r_2}$.  
It turns out that the kernel $\phi$ depends on 
$\bf{r_1}$ and $\bf{r_2}$ only through two variables 
$d_1=q_1r_{12}$ and $d_2=q_2r_{12}$, 
and can be expressed as 
\begin{eqnarray}
\phi(d_1,d_2)=\frac{2}{\pi^2} \int_0^{\infty}a^2da	
	      \int_0^{\infty}b^2db \ W(a,b)
		T(\nu(a),\nu(b),\nu^{'}(a),\nu^{'}(b)),
\end{eqnarray}
where $W$ and $T$ are defined as 
\begin{eqnarray}
W(a,b)&=&2[(3-a^2)b\cos b\sin a+
                        (3-b^2)a\cos a\sin b+	\nonumber	\\
			&& (a^2+b^2-3)\sin a\sin b-3ab\cos a\cos b]/a^3b^3, 	\\
T(w,x,y,z)&=&\frac{1}{2}[\frac{1}{w+x}+ \frac{1}{y+z}] \times
[\frac{1}{(w+y)(x+z)}+ \frac{1}{(w+z)(y+x)}]. 
\end{eqnarray}
The quantities $\nu$ and $\nu^{'}$ are given by
$\nu(u)=u^2/2h(u/d_1)$ and 
$\nu^{'}(u)=u^2/2h(u/d_2)$ with 
$h(t)=1-\exp (-4\pi t^2/9)$.

The universal function $q_0$ reads as 
\begin{equation}
q_0({\bf r})=-\frac{4\pi}{3}
\epsilon_{xc}^{LDA} n({\bf r}) - \frac{Z_{ab}}{9}
s^2({\bf r})
k_F({\bf r}).
\end{equation}
Here, the Fermi wave vector $k_F$ and the reduced gradient $s$  are  
\begin{equation}
k^3_F ({\bf r})=3\pi^2n({\bf r}), \ 
s({\bf r})=\frac{\nabla n({\bf r})}{
2k_F({\bf r})n({\bf r})},
\end{equation}
and 
$Z_{ab}=-0.8491$.

The nonlocal correlation energy in Eq. (\ref{eqnl}) is expressed as a double
spatial integral. To alleviate the $O(N^2)$ evaluation of such the integral, 
we adopt the algorithm by 
Rom\'{a}n-P\'{e}rez and Soler \cite{RS}, 
which transforms the double real space integral to reciprocal space 
and reduces the computational effort.

First, $\phi$ was interpolated as 
\begin{equation}
\phi(q_1,q_2,r_{12})=\sum_{\alpha,\beta}\phi(q_{\alpha},q_{\beta},r_{12})
p_{\alpha}(q_1) p_{\beta}(q_2), 	\label{eq10}
\end{equation}
where $q_{\alpha}$ are {\it fixed } values, chosen to ensure a good 
interpolation of function $\phi$. 
Here, we use cubic splines interpolation, in which $p_{\alpha}(q)$ 
is a succession of cubic polynomial in every interval 
$[q_{\beta},q_{\beta+1}]$, matching in value and the first two derivatives at 
every point $q_{\beta}$.

Substituting Eq. (\ref{eq10}) into (\ref{eqnl}),
\begin{equation}
E^{nl}_{c}= \frac{1}{2}  
\sum_{\alpha,\beta}
\int \int 
{ d{\bf r_1} d{ \bf r_2} }
\theta_{\alpha}({\bf r_1}) 
\theta_{\beta}({\bf r_2}) 
\phi_{\alpha \beta}(r_{12}), 
\end{equation}
with $\theta_{\alpha}({\bf r})=n({\bf r})p_{\alpha}(q_0({\bf r}))$ and 
$\phi_{\alpha \beta}(r_{12})\equiv\phi(q_{\alpha},q_{\beta},r_{12})$.
Now, with use of the convolution, just like the Coulomb energy, 
$E^{nl}_{c}$ becomes
\begin{equation}
E^{nl}_{c}= \frac{1}{2}  
\sum_{\alpha,\beta}
\int  { d{\bf k}}
\theta_{\alpha}^{*}({\bf k}) 
\theta_{\beta}({\bf k}) 
\phi_{\alpha \beta}(k), 
\end{equation}
where $\theta_{\alpha}({\bf k})$ and $\phi_{\alpha \beta}(k)$ are the 
corresponding Fourier transforms. In practice, 
 $\phi_{\alpha \beta}(k)$ was pre-calculated 
in spherical radial mesh of points $k$.
Then, $\phi_{\alpha \beta}(k)$ and its second derivative via 
cubic spline interpolation were stored for later use.
A logarithmic mesh of interpolation points $q_{\alpha}$, 
of which the total number is 20 in the present calculation, 
was used to describe $\phi$ up to a cutoff $q_c$ of 5.0 a.u..

\subsection{GGA charge density in the mixed-basis approach }
With one set of B-splines for the non-periodic $z$ 
direction and 2D plane waves for the periodic $xy$ plane,
the present mixed basis used to expand the wavefuction
is defined as
\begin{equation}
< {\bf r } | { \bf \ k_\parallel +  G_\parallel } ; j, \kappa > \
=
\frac{1}{\sqrt{A}}\
e^{i( {\bf k_\parallel +  G_\parallel} ) \cdot { \bf \rho } }
\ B_{j,\kappa}(z),
\end{equation}
where
$ {\bf G_\parallel} $
denotes an in-plane reciprocal lattice vector and   
$ {\bf k_\parallel} $ is the in-plane Bloch wave vector.
$A$ is the surface area of the system.
Therefore, the charge density can be written in the form  

\[
n({\bf r})  = \sum_{{\bf g}}  
\ n({\bf g},z)
\  e^{ i {\bf g} \cdot {\bf \rho} } \ ,
\]
where $ {\bf g}= {\bf G_\parallel}-{\bf G^{'}_\parallel}$.

Because the GGA energy functional 
depends upon $|\nabla n({\bf r})| $, the 
corresponding potential $v_{xc}$ is a functional of not only  
$|\nabla n| $, but also of 
$\nabla^2 n $ and  $\nabla n \cdot \nabla |\nabla n | $. In order to 
efficiently and precisely obtain $v_{xc}$,   
we used the method by White and Bird \cite{WB}. 
First of all, we interpolate 
$n({\bf g},z)$ along the $z$ direction by
using the Fourier interpolation technique:  
\[
n({\bf g},z)  = \sum_{g_z}  
\ n({\bf g},g_z)
\  e^{ i g_z z } \ .
\]
Then,  
\begin{equation}
n({\bf r})=\sum_{{ \bf G }}  n ( {\bf G}) 
e^{i {\bf G}\cdot { \bf r}},
\end{equation}
where ${ \bf G }$ is a compact notation for $({\bf g},g_z)$.

Following the procedure in Ref. \cite{WB}, 
\begin{equation}
\nabla n({\bf r})=\sum_{{ \bf G }} i { \bf G } n ( {\bf G}) 
e^{i {\bf G}\cdot { \bf r}}=
\frac{1}{N} 
\sum_{{ \bf G,R }} i { \bf G } n ( {\bf R}) 
e^{i {\bf G}\cdot ({ \bf r-R})} 
\end{equation}

\begin{equation}
=\frac{1}{N} 
\sum_{{ \bf G }} i { \bf G } 
\left( 
\sum_{{ \bf R }}n ( {\bf R}) 
e^{-i {\bf G}\cdot { \bf R}} 
\right) 
e^{ i {\bf G}\cdot { \bf r}}, 
\end{equation}
with $N$ real space points ${\bf R}$ of the fast Fourier-transform (FFT) grid set. 

We define $f_{xc}$ such that 
\begin{equation}
E_{xc}[n]=\int f_{xc}( n({\bf r}), |\nabla n({\bf r})|)d{\bf r}. 
\end{equation}
$E_{xc}[n]$ can be approximated by 
\begin{equation}
E_{xc}[n] \approx 
\frac{V}{N}\sum_{{ \bf R }} f_{xc}( n({\bf R}), |\nabla n({\bf R})|). 
\end{equation}
The associated xc potential at the FFT grid point {\bf R}
can be obtained efficiently through
\begin{eqnarray}
v_{xc}({\bf R}) &=& 
\frac{N}{V}\frac{d E_{xc}}{d n({\bf R})} 	\\
&=&\frac{\partial f_{xc}}{\partial n ({\bf R})}
+\sum_{ {\bf R^{'} }} \frac{\partial f_{xc}}{\partial \nabla n ({\bf R^{'}})}
\cdot \frac{d \nabla n ({\bf R^{'}})} {d n ({\bf R})}  \\
&=&\frac{\partial f_{xc}}{\partial n ({\bf R})}
+ \frac{1}{N} 
\sum_{{ \bf G,R^{'} }} i { \bf G } 
\cdot  
 \frac{\partial f_{xc}}{\partial \nabla n ({\bf R^{'}})}
e^{i {\bf G}\cdot ({ \bf R^{'}-R})} 	\\
&=&\frac{\partial f_{xc}}{\partial n ({\bf R})}
+ \frac{1}{N} 
\sum_{{ \bf G,R^{'} }} i { \bf G } 
\cdot  
\frac{\nabla n ({\bf R^{'}})}{n ({\bf R^{'}})}
 \frac{\partial f_{xc}}{\partial |\nabla n ({\bf R^{'}})|}
e^{i {\bf G}\cdot ({ \bf R^{'}-R})}. 	
\end{eqnarray}
Given the charge density on the FFT grid points, only eight FFT's are required
to obtain $v_{xc}$. That is computationally moderate with respect to 
the derivation of the
second derivative needed to evaluate the conventional potential via  
\begin{equation}
v_{xc}({\bf r})= \frac{\partial f_{xc}}{\partial n ({\bf r})}-
\nabla \cdot \frac{\partial f_{xc}}{\partial \nabla n ({\bf r})}.
\end{equation}

\subsection{vdW-DF total energy} \label{section}
For the vdW-DF total energy functional, we first performed the self-consistent
total energy calculation using the GGA-PBE functional \cite{PBE}. With the 
converged charge density obtained in the previous step,  
the revPBE exchange energy \cite{ZY} and LDA correlation energy \cite{PW}
 are evaluated and substituted for 
the GGA-PBE counterparts, and the nonlocal correlation energy $E^{nl}_{c}$ 
is added. Now, the vdW-DF energy functional is written 
\begin{equation}
E^{vdW-DF}= E^{PBE}- E^{PBE}_{xc}+  (
E^{revPBE}_{x}+
E^{LDA}_{c}+
E^{nl}_{c}  ). 
\end{equation}
The last three terms in the above equation are 
treated as a post-GGA perturbation because of their low sensitivity to 
the choice of GGA electronic density. By the present approach, $E^{nl}_{c}$ was obtained 
via 
\begin{equation}
E^{nl}_{c}= \frac{1}{2}A
\sum_{\alpha,\beta} \sum_{\bf g}
\int  { d{k_z}}
\theta_{\alpha}^{*}({\bf k})
\theta_{\beta}({\bf k})
\phi_{\alpha \beta}(k),
\end{equation}
with ${\bf k}=({\bf g},k_z)$.

The interplanar binding energy per surface atom $E_b$ is defined as 
\begin{equation}
E_b = (E_{bilayer}-2E_{graphene})/N_b
\end{equation}
where $E_{bilayer}$ and $E_{graphene}$  
are respectively the total energy of the bilayer graphene system and   
that of the system containing only one graphene sheet. 
$N_b$ is the number of atoms in {\it one single} graphene sheet.

\section{APPLICATIONS OF PRESENT METHOD} \label{cm}
To test the present approach, we apply it to investigate the
vdW interaction between graphene sheets. 
The calculations were carried out with
the unit cell containing two graphene sheets. 
Here, we study both AA and AB stacking, as depicted in Fig. \ref{fig1}. 
 In the former case,
the carbon atoms of the adjacent sheets are aligned directly on top 
of each other. In the latter case, 
the energetically more stable structure, the 
graphene layers are shifted relative to each other such that 
half of the atoms are located exactly over the center of a hexagon and 
another half lie directly on top of the atoms in the second graphene sheet. 

All C atoms in the graphene sheet were kept at the ideal positions. The
in-plane lattice constant $a_0$ was fixed to 
the experimental value of 2.461 \AA, while the inter-plane distance $d$ 
is allowed to vary. 
A FFT mesh with the grid spacing of  
0.08 \AA~for the charge density are chosen for accurate total
energy calculations. A mixed basis set with 
34 B-splines distributed over a maximum range of 6.0 $a_0$ 
and 2D plane waves with an energy cutoff of 30 Ry 
are used to expand the wavefunction.   
The $7 \times 7$ Monkhorst-Pack grids including $\Gamma$ point
were taken to sample the surface Brillouin zone. 
We used the Vanderbilt's ultra-soft pseudopotential (USPP) \cite{DV}.
The C USPP was generated from
the Vanderbilt's code \cite{vancode} and its quality was   
examined previously \cite{RHC}. 
The potential is determined self-consistently until its
change is less than $10^{-7}$ Ry. Finally, the vdW-DF 
total energy and the associated binding energy are calculated according to the 
procedure described in Sec. \ref{section}.
For comparison, we also performed calculations by using 
the standard supercell approach implemented in the 
popular VASP code with the projector-augmented-wave potential 
(PAW) \cite{KJ,KF}.  
A typical vacuum space of 10 \AA~ required in VASP was 
used in the calculation.

The binding energy of bilayer graphene in the AA stacking 
as a function of interlayer 
separation $d$ is shown in Fig. \ref{fig2}.  
The VASP counterpart is also plotted for comparison. 
Obviously, the GGA-PBE calculations show no relevant minimum for the graphite 
binding energy, reflecting the failure to include the proper long-range 
dispersive interaction within the GGA approximation. On the other hand, 
with the vdW-DF xc functional expressed in Eq. (\ref{vdw-xc}), 
we obtained for the graphene pair a binding energy 
of 47 meV/atom for the AA stacking. 
More importantly, the results obtained with our algorithm agree nicely with those by 
the popular VASP code. We are then convinced that the present program 
has been implemented successfully for the vdW interaction and the 
outcomes are very reliable. 

We also calculated for the AB stacking and    
the resulting binding energy is displayed in Fig. \ref{fig3}. 
Clearly, the AB stacking is  
energetically more stable than the AA stacking, in agreement with the 
experiment that natural graphite occurs mainly with AB stacking order \cite{THZ}. 
The binding energy for the AB stacking was found to be 50.5 meV/atom at the distance 
of 3.7 \AA~ and that for the less favored AA stacking was 47.0 meV/atom at the distance
of 3.8 \AA.  The results are consistent with data reported in the 
literature \cite{NPA},\cite{CSLL}-\cite{CKHS}. It can be seen from  
this figure that the $E_b$ curve for the AB stacking merges into that for the AA
stacking at large separation $d$, which seemingly indicates that 
the total energy will be less sensitive to the orientation of the two graphene
sheets if the separation is not too close.    

Actually, contrary to the covalent bond, 
the charge density distribution near the individual graphene sheet would not 
be noticeably affected via 
the weak vdW interaction from other sheets unless the sheet separation  
is close enough such that it begins to 
'contact' or even overlaps with the density from the adjacent sheet. In that case, 
the shape of the density distribution around the graphene sheet
will be distorted because of the dominant strong electron repulsion.   

To justify this assertion, we use
a rigid-density model, i.e., the whole charge density of the system in the AA stacking 
is simply assumed to be the sum of those self-consistently calculated 
for the individual single layer. Then we employed such charge density to
re-calculated the total energy $E^{tot,AA}_{rigid}$ and compared it 
to the total energy $E^{tot,AA}_{scf}$ with self-consistent density calculations. 
The results are summarized in Table \ref{tab2}. Here, we chose seven cases with 
various interlayer separation $d$, as also indicated in Fig. \ref{fig3}.
Clearly, the total energy per surface atom 
by the rigid-density model is very similar to the self-consistent total energy.  
The energy difference only becomes notable (2.7 meV/atom)  
for Case 7 with $d=2.95$ \AA, which already enters the electron-repulsive
region. With detailed analysis of energy components, we found that even 
though the energy difference from the kinetic energy part is somewhat 
sizable (for example, 32 meV/atom for Case 4 with $d=3.69$ \AA), but it is 
largely compensated for with the Hartree energy part, leading to 
almost the same value in total. 

When the charge density of the single graphene sheet 
was shifted a bit to the second one, we obtained similar conclusion, as 
shown in Table \ref{tab2} for the AB stacking. Therefore, it is reasonable to 
expect that if the first graphene sheet is rotated with respect to the 
second to become a Moir\'{e} rotated pattern, the rigid-density model can still hold to  
efficiently predict a reliable binding energy for such a 
large complex system. This finding is promising in searching for the true
ground-state atomic configuration for vdW-dominated graphene-based materials,
like MoS2/graphene heterostructures \cite{PHABNP}. Instead of using the  
very time-consuming self-consistent approach for such materials, 
the rigid-density model allows us to only focus on accurate charge density calculations
of every individual slab of different type.  
%
\section{CONCLUSIONS} \label{con}
In conclusion, we have successfully implemented
the van der Waals (vdW) density functional proposed by Dion {\it et al.} \cite{DRSLL}
in our mixed-basis approach for investigating the bi-layer graphene system. 
As compared to the conventional supercell model with
alternating slab and vacuum regions, it is a real space approach along
the non-periodic direction. Therefore, the number of the 
basis functions used to expand the wavefunction is significantly reduced, 
especially for large complex systems. 

We also found that the self-consistent total energy 
obtained for the bilayer 
system is not significantly different from that with charge density 
assumed to be the simple 
sum of those for the two individual single-layer system, except when the
distance between the two layers is close enough that the  
strong electron-repulsion dominates. Such observations help us to propose a
rigid-density model which can efficiently 
calculate the binding of vdW-dominated 2D systems with  Moir\'{e} pattern configurations.  
\begin{acknowledgments}
This work was supported by Ministry of Science and Technology under 
grant numbers MOST 106-2112-M-017 -003 and MOST 106-2112-M-001-022 
and by National Center for Theoretical Sciences of Taiwan.
\end{acknowledgments}

\appendix* 
\section{}
\newpage
\begin{center}
\large {\bf FIGURE CAPTIONS}    \normalsize
\end{center}
Fig. 1: Atomic structure of the graphite in both AA and AB stacking.  \\ \\
Fig. 2: (Color online) Binding energy for the two-layer graphenes in AA stacking 
        calculated with
        GGA-PBE and vdW-DF xc functionals. The counterparts obtained by VASP are 
	also displayed for comparison.  \\ \\
Fig. 3: (Color online) Binding energy for the two-layer graphenes in both AA and 
        AB stacking calculated with vdW-DF xc functional.  \\ \\
\newpage
\begin{figure}[h]
        \caption{ } \label{fig1}
\includegraphics{fig1.eps}
\end{figure}
\newpage
\begin{figure}[h]
        \caption{ } \label{fig2}
\includegraphics{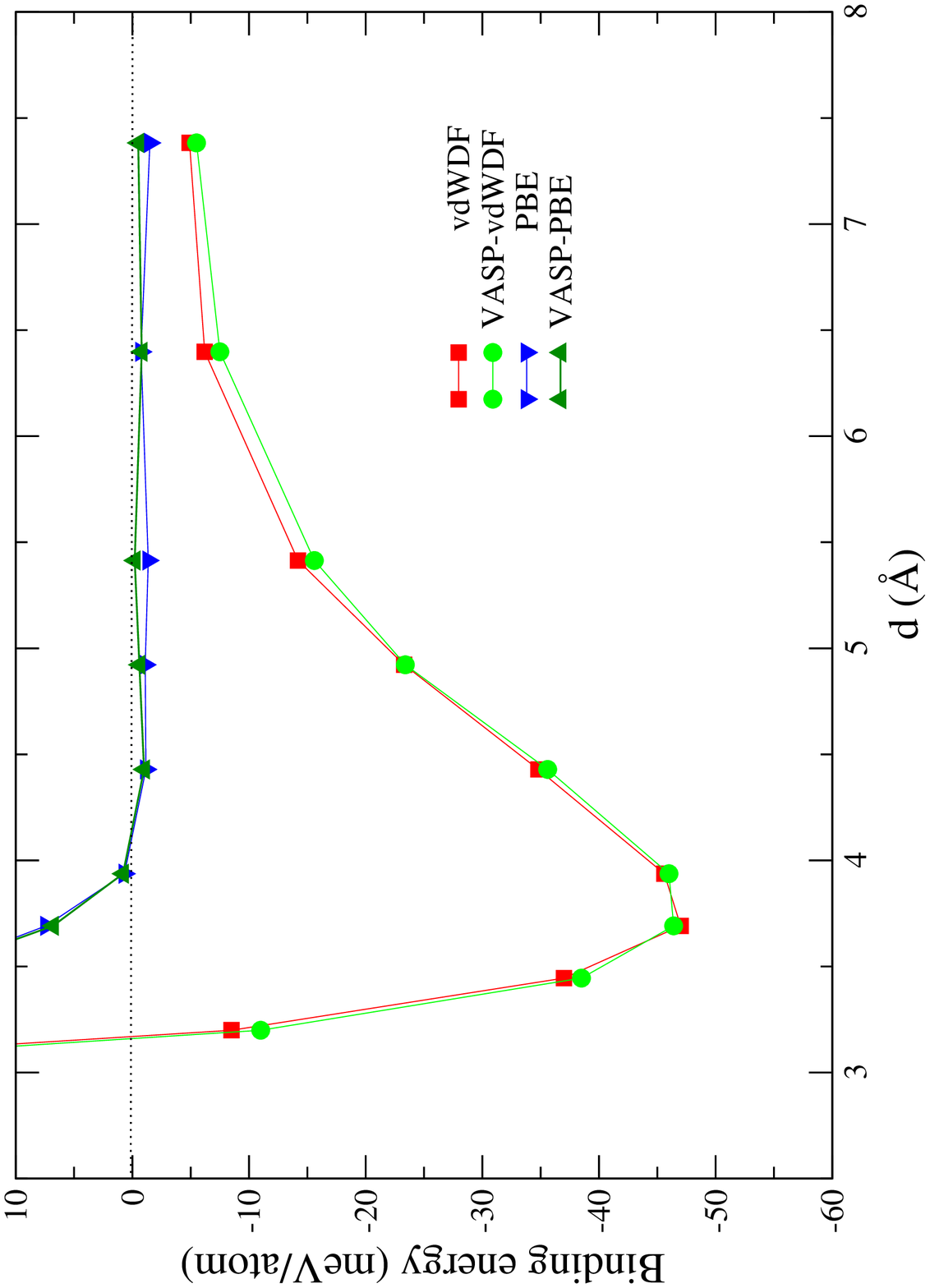}
\end{figure}
\newpage
\begin{figure}[h]
        \caption{ } \label{fig3}
\includegraphics{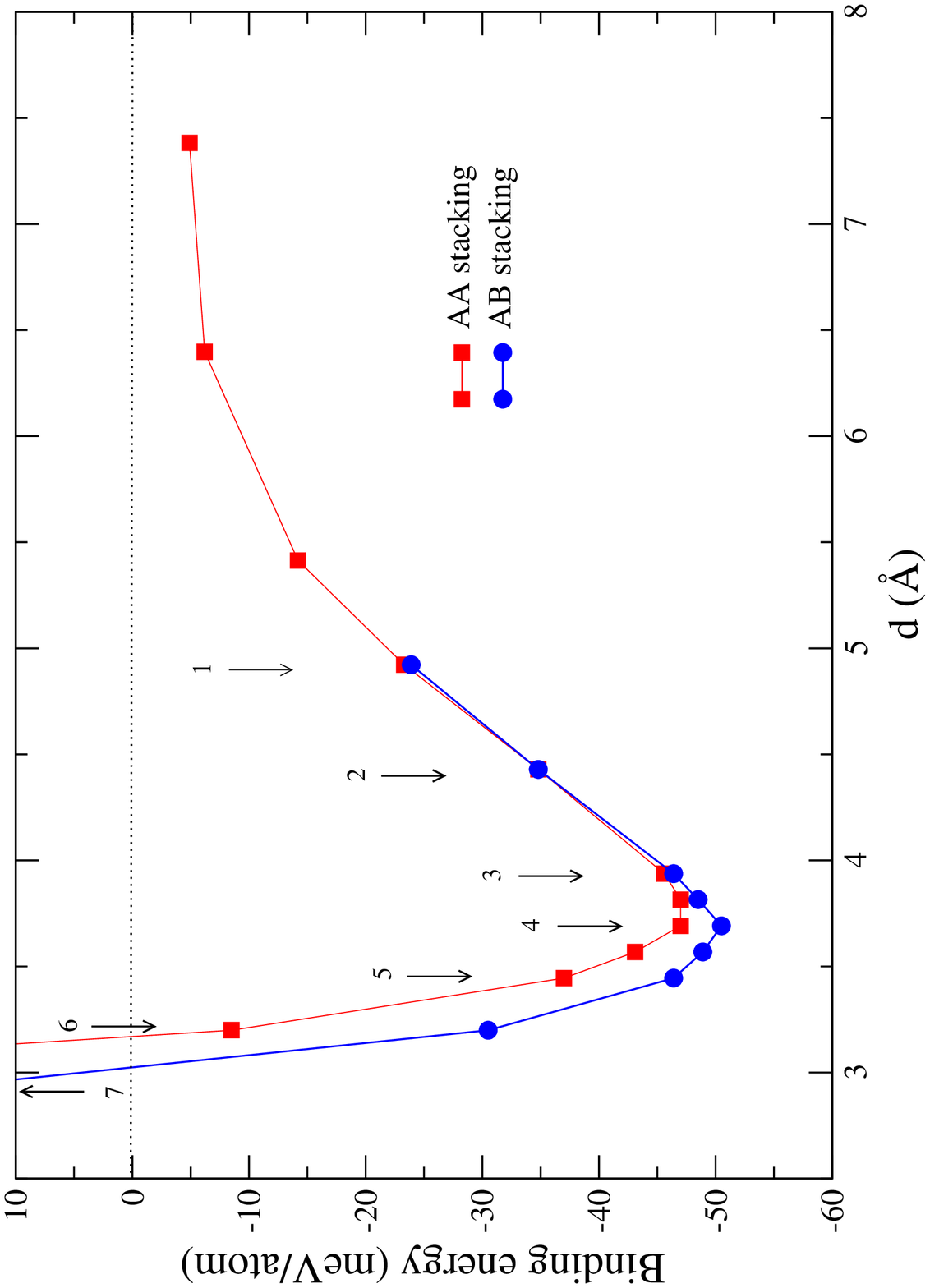}
\end{figure}
\newpage
\begin{table}[t]
        \caption{ Total energy difference per surface atom between those 
		  obtained by the self-consistent calculation and 
		  rigid-density model for the bilayer graphene system with  
	          various interlayer separation $d$ in both AA and AB stacking. 
		  See text for details.}  \label{tab2}
        \normalsize
        \begin{center}
                \begin{tabular}{llcc}
\hline  \hline
case No.  \hspace{10mm} & $d$    \hspace{10mm}       & $E^{tot,AA}_{scf}- E^{tot,AA}_{rigid}$         
     \hspace{10mm}   & $E^{tot,AB}_{scf}- E^{tot,AB}_{rigid}$        \\
\hspace{10mm} & (\AA) \hspace{10mm}           & (meV/atom) \hspace{10mm}        & (meV/atom)        \\
\hline
1 &4.92              &0.2            &0.1           \\
2 &4.43              &-0.1           &-0.1            \\
3 &3.94              &0.2            &-0.3       \\
4 &3.69              &0.1            &0.2  	\\
5 &3.45              &-0.4           &0.2  	\\
6 &3.20              &0.7            &0.4  	\\
7 &2.95              &2.7            &1.1  	\\
\hline \hline
                \end{tabular}
        \end{center}
\end{table}

\begin{thebibliography}{99}
\bibitem{BKLW}
M. Breitholtz, T. Kihlgren, S. -\AA. Lindgren, and L. Walld\'{e}n, 
Phys. Rev. B {\bf 66}, 153401 (2002).
\bibitem{ZKSH}
E. Ziambaras, J. Kleis, E. Schr\"{o}der, P. Hyldgaard,
Phys. Rev. B {\bf 76}, 155425 (2007).
\bibitem{PHABNP}
D. Pierucci,H. Henck, J. Avila, A. Balan, C. H. Naylor, G. Patriarche,
Y. J. Dappe, M. G. Silly, F. Sirotti, A. T. C. Johnson, M. C. Asensio,
and A. Ouerghi,
Nano Lett. {\bf 16}, 4054 (2016). 
\bibitem{RJHSLL}
H. Rydberg, N. Jacobson, P. Hyldgaard, S. I. Simak, B. I. Lundqvist, and 
D. C. Langreth,
Surf. Sci. {\bf 532-535}, 606 (2003). 
\bibitem{LM}
I. -H. Lee and R. M. Martin, 
Phys. Rev. B {\bf 56}, 7197 (1997).
\bibitem{B}
J. C. Boettger, Phys. Rev. B {\bf 55}, 11202 (1997).
\bibitem{NPA}
D. Nabok, P. Puschnig, and C. Ambrosch-Draxl, 
Comp. Phys. Comm. {\bf 182}, 1657 (2011).
\bibitem{DRSLL}
M. Dion, H. Rydberg, E. Schr\"{o}der, D. C. Langreth, and B. I. Lundqvist,
Phys. Rev. Lett. {\bf 92}, 246401 (2004), {\bf 95}, 109902(E) (2005). 
\bibitem{RHC}
C. Y. Ren, C. S. Hsue and Y.-C. Chang, Comp. Phys. Comm. {\bf 188}, 94 (2015).
\bibitem{RCH}
C. Y. Ren, Y.-C. Chang, and C. S. Hsue, Comp. Phys. Comm. {\bf 202}, 188 (2016).
\bibitem{deBoor}
Carl deBoor, {\it A practical Guide to Splines}, (Springer, New York, 1987).
\bibitem{JBS}
W. R. Johnson, S. A. Blundell, and J. Sapirstein, Phys. Rev. A {\bf 37}, 307 (1988).
\bibitem{JH}
H. T. Jeng, and C. S. Hsue, Phys. Rev. B {\bf 62}, 9876 (2000).
\bibitem{RJH}
C. Y. Ren, H. T. Jeng, and C. S. Hsue, Phys. Rev. B {\bf 66}, 125105 (2002).
\bibitem{LC}
G.-W. Li and Y.-C. Chang, Phys. Rev. B {\bf 48}, 12032 (1993).
\bibitem{LC1}
G.-W. Li and Y.-C. Chang, Phys. Rev. B {\bf 50}, 8675 (1994).
\bibitem{ZY}
Y. Zhang and W. Yang, Phys. Rev. Lett. {\bf 80}, 890 (1998). 
\bibitem{PW}
 J. P. Perdew and Y. Wang, Phys. Rev. B {\bf 45}, 13244 (1992) and references therein.
\bibitem{RS}
G. Rom\'{a}n-P\'{e}rez, and J. M. Soler, Phys. Rev. Lett. {\bf 103}, 096102 (2009). 
\bibitem{WB}
J. A. White and D. M. Bird, Phys. Rev. B {\bf 50}, 4954(R) (1994).
\bibitem{PBE}
J. P. Perdew, K. Burke, and  M. Ernzerhof, Phys. Rev. Lett. {\bf 77}, 3865 (1996). 
\bibitem{DV}
D. Vanderbilt, Phys. Rev. B {\bf 41}, 7982 (1990).
\bibitem{vancode}
http://www.physics.rutgers.edu/~dhv/uspp/.
\bibitem{KJ}
G. Kresse and D. Joubert, Phys. Rev. B {\bf 59}, 1758 (1999).
\bibitem{KF}
G. Kresse and J. Furthm\"{u}ller, Comput. Mater. Sci.  {\bf 6}, 15 (1996).
\bibitem{THZ}
P. H. Tan, W. P. Han, W. J. Zhao, Z. H. Wu, K. Chang, H. Wang, Y. F. Wang,
N. Bonini, N. Marzari, N. Pugno, G. Savini, A. Lombardo, A. C. Ferrari,
Nat. Mater. {\bf 11}, 294 (2012).
\bibitem{CSLL}
S. D. Chakarova-K\"{a}ck, E. Schr\"{o}der, B.I. Lundqvist, and D.C. Langreth, 
Phys. Rev. Lett. {\bf 96}, 146107 (2006). 
\bibitem{CKHS}
S. D. Chakarova-K\"{a}ck, J. Kleis, P. Hyldgaard, and E. Schr\"{o}der, 
New J. Phys. {\bf 12}, 013017 (2010). 
\end{thebibliography}
\end{document}